\documentclass{sig-alternate}
\toappear{To appear in the Proceedings of the 2014 International Conference on Software Engineering, track New Ideas and Emerging Results (ICSE NIER 2014)}

\usepackage[T1]{fontenc}
\usepackage{balance}
\usepackage{numprint}
\usepackage{framed}
\usepackage{tabularx} 
\usepackage{listings} 
\usepackage{paralist}
\usepackage{enumitem}
\usepackage{nohyperref,url}
\usepackage{todonotes}
\usepackage{listings} 
\usepackage{multicol}
\usepackage{datatool}
\usepackage{multirow}
\usepackage{booktabs}
\usepackage{rotating}

\npthousandsep{,}
\npdecimalsign{.}
\title{Do the Fix Ingredients Already Exist? \\ An Empirical Inquiry into the Redundancy Assumptions of Program Repair Approaches}

\author{Matias Martinez$^\dagger$\hspace{1cm} Westley Weimer$^\ddagger$ \hspace{1cm} Martin Monperrus$^\dagger$\\
\affaddr $\dagger$ University of Lille \& INRIA, France \hspace{1cm}   $\ddagger$  University of Virginia, USA\\
}

\begin{document}
\conferenceinfo{ICSE}{'14, May 31 - June 7, 2014, Hyderabad, India}
\CopyrightYear{14}
\crdata{978-1-4503-2768-8/14/05}
\maketitle

\begin{abstract}

Much initial research on automatic program repair has focused on experimental results to probe
their potential to find patches and reduce development
effort. Relatively less effort has been put into understanding the hows and
whys of such approaches. For example, a critical assumption of the
GenProg technique is that certain bugs can be fixed by copying and
re-arranging existing code. In other words, GenProg assumes that the fix
ingredients already exist elsewhere in the code. In this paper, we
formalize these assumptions around the concept of ``temporal redundancy''.
A temporally redundant commit is only composed of what has already existed
in previous commits. 
Our experiments show that a large proportion of commits that add existing code are temporally redundant. 
This validates the fundamental redundancy assumption of GenProg.
\vspace{-0.3cm}
\end{abstract}

\category{D.2.5}{Software Engineering}{Testing and Debugging}
\terms{Experimentation}
\keywords{Automatic program fixing; automatic software repair; mining software repositories}


\section{Introduction}


%

To some extent, each program repair technique is based on an underlying assumption. GenProg's
``secret sauce'' ~\cite{Weimer2009,LeGoues2012}  is the assumption that large programs contain the seeds of their own repair and thus that rearrangements of existing statements
can fix most bugs. This redundancy assumption is also behind four out of PAR's ten repair templates ~\cite{Kim2013}.
By contrast, SemFix~\cite{nguyen13} makes different assumptions: it is based on the idea that some bugs can be repaired by changing only one variable assignment or if conditional expression.
Although extensive empirical evaluations of the bug fixing
powers and costs of these techniques have been carried out, less work has
focused on exploring or validating such fundamental assumptions. 

In this paper we focus on the assumptions made by techniques like PAR and
GenProg, and ask whether it makes sense to re-arrange existing code or
code changes to fix bugs. We formalize our investigation by defining a
concept of ``software temporal redundancy''. A commit is said to be
temporally redundant if it is only a re-arrangement of the code that has
already existed in previous commits. We provide further insight into the
problem by breaking down existing code along multiple dimensions, including
the granularity at which it is considered (e.g., line-level vs.
token-level) and the locality at which it is found (e.g., in the same file
vs. anywhere in the program). 

We measure temporal redundancy on a dataset of six open-source Java applications.
For instance, at token level, we find that 31-52\% of commits are temporally redundant. 
%
Our results show that a substantial fraction of software evolution takes
place in a ``closed world'', with no creation of new material. To our
knowledge, this paper is the first empirical quantification linking such
evidence to automated program repair. 

Our experiments enable us to understand the foundational assumptions of program repair approaches based on redundancy such as GenProg or PAR.
First, we validate the redundancy assumption: it \emph{does} makes sense to
fix defects by rearranging existing statements. 
Second, we find that 
different assumptions about how patches should be formed are significant:  
depending on the way one picks code from elsewhere in the program, the
search space is different in terms of size and potential for
success. Those results are directly actionable for improving such automated
program repair techniques. For example, our results indicate that only
considering local redundancy at the line level would decrease the repair
time without a similar reduction in success potential.


To sum up, our contributions are:
\begin{itemize}[itemsep=0.01pt]
\item the definition and formalization of software temporal redundancy;
\item an experimental protocol to measure software temporal redundancy using versioning data;
\item an empirical validation of the redundancy assumption of some program repair approaches on 16071 commits of open-source software applications;
\item an empirical analysis on the relation between the way existing
redundant code is selected for use in patches and the structure of the
underlying search space.
\end{itemize}

%


\section{Software Temporal Redundancy}\label{chap:temporal_redundancy}

Research in software evolution uses time as a primary focus. 
For most software projects, software evolution artifacts are captured by
version control systems (e.g. CVS, SVN or GIT).
In this section, we present a new commit-level software evolution metric called ``temporal redundancy''. 

\paragraph{Commits}

A \emph{commit} in a version control system consists of new file versions.
Conceptually, a commit can be viewed in two ways: as a set of file pairs (before and after the commit),
or as a set of changes (applied to the version before the commit to obtain
the version after it). 

In this paper, we use this change-based view of commits. We consider that a commit adds and/or deletes new source code fragments. 
An \emph{update} is considered as the combination of a deletion and an addition.
We do not consider commits done on other artifacts than source code.

\paragraph{Fragment Redundancy}

We use \emph{fragment} to denote a substring of source code. 
For instance, the source code line ``for (int i=0;i<n;i++)'' is a fragment.
Fragments are always defined according to a level of granularity.
In this paper we consider two different levels:
lines (as separated by line breaks) and tokens (as separated by lexing
rules). 

A fragment $F$ is \emph{snapshot redundant} at time $T$ if another instance of that same fragment $F$ exists elsewhere in the program at time $T$. This is the redundancy studied by Gabel and Su \cite{gabel2010study} and used by GenProg~\cite{Weimer2009}. In this paper we consider a richer notion of redundancy that includes historical context.

A fragment $F$ is \emph{temporally redundant} at time $T$ if that same
fragment $F$ has already been seen during the history of the software under
analysis (i.e., at time $T' < T$). For instance, literal ``42'' might
be added in version \#1, be removed in version \#2 and reused again in
version \#3. In the commit of version \#3, the ``42'' fragment is temporally
redundant. Thus, in this paper, once a fragment has appeared it is always
subsequently viewed as a potential source of redundancy. We consider the
first version of a program to be created by insertions from an empty
initial program. 


\paragraph{Commit Redundancy}\label{chap:commit_redundancy}

A source code commit is composed of added and/or deleted fragments.
Using a cooking metaphor, the added fragments are the ``ingredients'' of the commit. 

We define a \emph{temporally redundant commit} as a commit for which all added fragments are (individually) temporally redundant.
More formally, we define a commit $C_j$ performed at the time $T_j$ as a
set $S_j$ of added fragments and a set $R_j$ of removed fragments.
Let $C$ be the set of all commits for a program. Then 
a temporally redundant commit $C_j$ satisfies
\[
\forall f \in S_j| ~ \exists C_i \in C| ~ 
T_i < T_j \wedge f \in S_i 
\]

For such commits, no new fragments are invented and no fresh material
is introduced: the commit is only a re-arrangement of insertions that
have already been seen in previous commits. 


%
%
%

\paragraph{Scope of Temporal Redundancy}
\label{chap:redundancy_scope}
A fragment is temporally redundant if that same fragment appeared
in a previous commit.  This kind of redundancy has a \emph{global scope}:
the location of the previous fragment instance does not matter.

We now define a more restricted \emph{local scope} notion 
of temporal redundancy. A fragment is locally temporally redundant
if that same fragment has been used in a previous commit to the same
\emph{file}. 

For example, consider two files, $F_1$ and $F_2$, each containing 3
fragments:
$ F_1= \{a,b,c\}$, $F_2= \{d,e,f\}$.
Suppose commit $C_1$ adds fragment $c$ to file $F_2$. 
For that commit, fragment $c$ 
is global temporally redundant (already available in $F_1$), but not 
locally temporally redundant (never previously available in $F_2$).
Suppose commit $C_2$ introduces another version of $F_2$ replacing fragment
$e$ with $d$. In that commit, fragment $d$ is locally redundant
(since $d$ was previously available in $F_2$).



\section{Measuring  Redundancy}
\label{chap:measuring_temporal_redundancy}

We now present an experimental design to measure the fraction of relevant
commits to a program that are temporally redundant. 
\paragraph{Experimental Protocol}

Given a level of granularity and a scoping level, 
our experimental protocol to measure the temporal redundancy of the
evolution of a program consists of the following phases:

\begin{inparaenum}[\itshape a\upshape)]
\item Retrieving commits.   
All commits of the program under analysis are collected from the 
repository. 
 
\item Filtering commit files. Only commits to executable code 
are retained; commits to test cases are discarded. 

\item Fragmenting commits.
We split each relevant file into fragments at a given level of
granularity (e.g., lines or tokens). 
This results in a before-commit and an after-commit sequence of fragments. 
We use the Myers differencing algorithm~\cite{Myers86} to compare both
fragment sequences and obtain the added fragments of the commit. 

\item Filtering fragments.
We filter out whitespace and comments. In this paper we are only interested
in the evolution of executable code and not in indentation or
documentation.

\item Selecting acceptable commits.
We select those commits that introduce at least one fragment after
filtering. We call such commits \emph{acceptable}.

\item Indexing fragments. 
We consider each added fragment in each acceptable commit in ascending
temporal order. If a fragment has not been encountered previously at the
given scoping level (i.e., global or local), we index
it with the date of its first introduction. 

\item Measuring temporal redundancy.  
The \emph{temporal redundancy} of the entire program's evolution
is the fraction of acceptable commits that are temporally redundant (see
Section~\ref{chap:commit_redundancy}). 
For example, a temporal redundancy of 0.1 means that 10\% of acceptable
commits are temporally redundant commits.

\end{inparaenum}

\paragraph{Dataset}
We use six open-source Java projects to measure the temporal redundancy.
They are: 
Apache Log4j, JUnit, Picocontainer, Apache Commons Collections, Apache Commons Math, and Apache Commons Lang.
The inclusion criterion is as follows: the Apache projects were used in previous
research on automatic program repair~\cite{Kim2013}, while the remaining
two are Java projects mentioned in previous research on software evolution \cite{kagdi2007survey}.  
After applying the filters presented above on the 16071 commits of the dataset, we obtain 7076 acceptable commits.

\newcommand{\head}[1]{ \parbox[t]{0.6in}{\scriptsize{#1}}}
\newcommand{\headsm}[1]{ \parbox[t]{0.25in}{\scriptsize{#1}}}
\newcommand{\col}[1]{ \bf{\tiny{C #1}}}
\newcolumntype{s}{>{\hsize=4.0\hsize}X}
\begin{table*}[t]
\centering
\caption{The temporal redundancy of six open-source applications.}
\begin{tabular}{| l |r ||  r |  r | r | r || r |  r | r | r |} 
\hline
 \multirow{3}{*}{Program}& \multirow{3}{*}{\head{Acceptable Commits}}& 
 \multicolumn{4}{c ||}{Line granularity }&\multicolumn{4}{c|}{Token granularity}\\

\cline{3-10}
&& \multicolumn{2}{c|}{Global}& \multicolumn{2}{c||}{Local}&\multicolumn{2}{c|}{Global}& \multicolumn{2}{c|}{Local}\\
\cline{3-10}

&
&\head{Temporal  redundancy}&\headsm{Pool Size}
&\head{Temporal redundancy}&\headsm{Pool Size}
&\head{Temporal redundancy}&\headsm{Pool Size}
&\head{Temporal redundancy}&\headsm{Pool Size}\\
\hline

log4	
        &1687
        &\bf{9\%} &43313  
        &\bf{6\%} & 57 
        &\bf{39\%}& 14294 
        &\bf{19\%}& 71 \\ 

junit   
        &713
        &\bf{17\%} & 8855
        &\bf{16\%} &18
        &\bf{43\%} & 3256
        &\bf{29\% }& 72.5\\

pico    
        &157 
        &\bf{3\%} & 16911
        &\bf{2\%} & 22.5
        &\bf{31\%}& 6273
        &\bf{8\%} &46\\

collections
        &1019
        &\bf{7\%}&25406
        &\bf{4\%}&35
        &\bf{52\%}&4163
        &\bf{23\%}&85.5\\

math    
        &2210
        &\bf{6\%}&69943
        &\bf{4\%}&37
        &\bf{45\%}&20742
        &\bf{18\%}&100.5\\

lang    
        &1290
        &\bf{8\%}&22330
        &\bf{6\%}&63
        &\bf{50\%}&6692
        &\bf{29\%}&98\\
\hline
\col{1}&\col 2&\col 3&\col 4&\col 5&\col 6&\col 7&\col 8&\col 9&\col {10}\\
\hline

\end{tabular}
\label{tab:scope-redundancy-general-analysis}
\end{table*}

\section{Empirical Results}  
\label{sec:results}

We now present our empirical results on the temporal redundancy of software
(as defined in Section~\ref{chap:commit_redundancy}) following the
experimental design presented in
Section~\ref{chap:measuring_temporal_redundancy}.

\subsection{Line-Level Temporal Redundancy}
\label{chap:line_experiment}
\emph{Research question: What is the amount of line-based temporal redundancy?}

For each application of our dataset, we measured
the total number of acceptable commits within the analysis timespan (the complete data is available at \url{http://goo.gl/k0rZWc})
and the global-scope line-level temporal redundancy.
Columns \#2 and \#3 of Table
\ref{tab:scope-redundancy-general-analysis} report the results.

For instance, for log4j (the first row), there are 1687 commits 
which add at least one executable line. 
Only 9\% of those 1681 are temporally redundant commits.

\emph{Overall, at the level of lines, 3--17\% of the accepted commits are
temporally redundant commits. Their basic ingredients are
only previously-inserted code.}

This has additional implications for automatic repair and code synthesis: 
for synthesizing those commits, the search-space has a finite number of atomic
building blocks (previously-observed line-level fragments). 
Theoretically, a redundancy-based approach should be able to synthesize all the temporally-redundant commits.

The interval  3--17\% is large and the reasons behind this variation are not obvious. 
One reason could be that the commit conventions used by the developers of a project  are different.  
For instance, some projects prefer to have small and atomic commits (one bug fix or feature per commit). 
Other projects are less restrictive on this point.
This has a direct impact on the redundancy: small and atomic commits are more likely to be redundant.
A deeper analysis of this question is future work.

\subsection{Token-Level Temporal Redundancy} 
\label{chap:granularity_experiment}
\emph{Research question: Is there a difference between line-based and token-based temporal redundancy?}

Before answering this question, we note that, analytically, all
temporally redundant commits at the level of lines are necessarily temporally redundant
commits at the level of tokens. Furthermore, a unique new line might be
exclusively composed of existing tokens. 
Consequently, the token-based temporal redundancy must be equal to 
or greater than line-based temporal redundancy.
We now measure the temporal redundancy at the line and token level.

Table \ref{tab:scope-redundancy-general-analysis} reports global scope
line-level (column \#3) and token-level temporal redundancy
(column \#7).
For instance, in log4j, there is a line-level temporal redundancy of 9\%
but a token-level redundancy of 39\%. 
Overall, at the token level, 29--52\% of commits are temporally redundant. 

For all projects, token-based temporal redundancy exceeds
line-based.
This follows from the analytical argument above and gives confidence 
in the experiment's construct validity. 

\emph{Overall, token-level temporal redundancy (between 29\% and 52\%) is much higher than the line-level temporal redundancy.}

This is very good news. For automated repair and code synthesis, 
a high temporal redundancy implies a smaller search space.
This holds for both the line and token level of granularity.
Our token-level temporal redundancy measurements imply that for between 
29\% and 52\% of accepted commits, synthesis and repair need never invent a
new token. 
For instance, repair or synthesis of arithmetic code need only consider
recombining existing literals and operations for one-third to one-half of
commits. 

We thus propose that the repair search space is composed of two
components: 
the search space of fragments (the atomic building blocks) and
the search space of their combinations.  Our experiment enables us to
precisely measure the former: we can count the total number of fragments
seen up to a given point in time. We call this a \emph{fragment pool}.
There is one fragment pool per level of granularity.  This is 
conceptually similar to the
pool of four DNA bases or the pool of amino acids in biology.

Table \ref{tab:scope-redundancy-general-analysis} gives the size of the
global scope fragment pool for line-level (column \#4) and token-level
(column \#8) analyses.
For instance, in the considered slice of history of log4j, there are 43313
different lines and 14294 different tokens that are involved in the
software evolution.
For all applications under study, the token pool at the point in time of
the last commit is much smaller than the line pool.

For automated repair or code synthesis, there is a tension between working
with the line pool or the token pool.  To some extent, the temporally redundant
commits correspond to the number of commits that can be synthesized.  With
the line pool, the combination of lines is much smaller (the combination
space is smaller) but fewer commits can be synthesized ($\sim$10\%).  With the
token pool, more commits can be synthesized ($\sim$40\%), but at the price of
exploring a much bigger combination space.



\subsection{Redundancy Scope Experiment}
\label{chap:scope_experiment}

\emph{Research question: Do file scope restrictions impact the amount of temporal redundancy?}

We now measure the temporal redundancy available at the local scope in
the same file (as defined in Section  \ref{chap:redundancy_scope}). 

Table \ref{tab:scope-redundancy-general-analysis} reports 
global and local scope temporal redundancy in our dataset.
For each granularity, there is one column ``Global'' and one column ``Local'' corresponding to the different scope. 
For instance, at the line level,
column \#3 is the temporal redundancy at the global scope and column \#5 gives it when considering a local scope.

As discussed in Section \ref{chap:line_experiment}, at the line granularity, there are between 3\% and 17\% of temporally redundant commits at the global scope.
At the local scope, there are between 2\% and 16\%.

The temporal redundancy of both scopes is of the same order of magnitude.
In all projects, more than half of the temporally redundant commits
actually have local temporal redundancy. Consequently, \emph{at line
granularity, most of the temporal redundancy is localized in the same
file}.

At token-level granularity, the results are similar: we find a large
amount of token-level local-scope temporal redundancy.  
Tokens are likely to be reused in commits impacting the same file.
This further indicates that the fragment locality matters during software evolution.
We note that the difference of redundancy between
global and local scopes is slightly higher at the token level (col. \#7 vs \#9) than at the line level  (col. \#3 vs \#5).  

This is again a promising result with respect to the search space of
automatic repair or synthesis. 
First, at the line level, the local scope pool is able to seed the same order of magnitude of commits as the global one.
In other words, it is almost as fertile as the global pool.
Second,
when one considers the local scope pool, the search space is much
smaller.  For instance, for log4j, the median local pool size at the line
level is 57 lines, compared to 43313 at the global scope level.
Restricting attention to the
local scope reduces the search space greatly while still enabling the
synthesis of a large number of commits. 
Those results are directly actionable for improving GenProg and other redundancy-based approaches: our results indicate that only considering local redundancy would decrease the repair time while keeping a high repair success potential.
This is in line with our body of previous experiments.


\section{Related Work}
\label{sec:rw}
Some work on code clone devetion also focuses on software redundancy,
studying the differences between line- and token-level
granularity~\cite{Kamiya2002}. 
Kim \emph{et al.}~\cite{Kim2005clone} considered code clones via a 
temporal perspective, linking clones together across versions.
By contrast, we study whether changes only contain existing code.
Most importantly, a change composed of many redundant fragments may not be
a clone in itself: each redundant fragment could come from different,
unrelated, locations or appear in a rearranged order. 

In previous work~\cite{Martinez2013},  we discussed the topology of the
repair space and efficient manners to navigate it. In this paper, we focus
on the relation between commits and the repair space, which has not been
previously addressed. 
 
Gabel and Su~\cite{gabel2010study} studied the uniqueness of
source code.  Their ``syntactic redundancy'' calculates the degree to which
portions of software applications are redundant.
Hindle and colleagues~\cite{hindle2012naturalness} studied the 
repetitiveness and predictability of code.  
Both consider software redundancy from a spatial viewpoint.
By contrast, we study temporal redundancy. Our measures are not present in,
and cannot be inferred from, their experiments. 


Nguyen \emph{et al.}~\cite{Nguyen-etal-13} presented a study of
repetitiveness of code changes in software evolution.
Where they measure the repetitiveness of code changes by abstracting over
commits and literals, we measure the redundancy of code changes at the
commit level considering real, non-abstracted code (e.g., we care about the
actual values of literals).

%

%

%

%

\section{Conclusion}
\label{cap:conclusion}
In this paper, we define a notion of software temporal redundancy and
empirically measure its presence in source code. 
For example, as many as
52\% of commits are composed entirely of previously-existing tokens. 
Our results have direct application to automated program repair. For
example, we find that searches for line-level repairs can focus on a single
file, dramatically reducing the search space without substantially reducing
repair success potential. 

Our future work will focus on measuring temporal redundancy on bug fix commits and on studying different granularities beyond lines and tokens (e.g. at the AST level, focusing on some kind of source code elements).
\section{Acknowledgments} 
This research is done with support from EU Project
Diversify FP7-ICT-2011-9 \#600654 and the NSF 
(SHF-0905236, CCF-1116289, CCF-0954024), AFOSR 
(FA9550-07-1-0532, FA9550-10-1-0277), and DARPA
(P-1070-113237). We thank Maxence G. de Montauzan and Sylvain Magnier for their participation to the experimentation.

\bibliographystyle{abbrv}
\bibliography{biblio-software-repair}

\end{document}